\documentstyle[aps]{revtex}
\begin{document}
\author{O.V.Zhirov
\footnote{e-mail address zhirov@inp.nsk.su}}
\address{Budker Institite of Nuclear Physics, 630090,Novosibirsk, Russia}
\title{Density-Density Correlators at finite separations in Infinite Banded Random Matrices.}
\date{\today}
\maketitle

\begin{abstract}
Using the BRM theory developed recently by Fyodorov and Mirlin we calculate
the density-density correlators for Banded Random Matrix of infinite size.
Within the accuracy of $1/b^2$ ($b$ is the matrix bandwidth) it appears 
to be the same in both cases of the orthogonal and unitary symmetry. Moreover, 
its form coincides exactly with the formula obtained long ago by Gogolin for  the
electron density-density correlator in strictly 1D disordered metals.

In addition to the ``fixed energy'' density-density correlator considered
in the solid state physics we calculate also the ``time averaged''  one,
which has different properties at small separations.  Our predictions are  
compared with the existing numerical  data.
\end{abstract}

\section{Introduction}

The Band Random Matrix(BRM) model is a good basis%
\footnote{%
See, e.g., \cite{Izra90,Fyod94,Izra97} as well as a recent very extensive
review \cite{Guhr97}, and references therein.}
for  understanding basic features of quantum chaotic systems. It looks even 
more attractive in view of its analytical solution obtained by Fyodorov and 
Mirlin \cite{Fyod91} in the framework of the Efetov's supersymmetry approach
\cite{Efet83}. Being inspired by numerous earlier attempts\cite{Casa90}
to understand scaling properties of the BRM model, the solution is proved to
be an adequate tool to address the problem.

Recently, further extensive numerical studies of the RBM model for the real  
symmetric matrices (we refer to this case as that of orthogonal ensemble) 
has been done in the work\cite{Izra97}. 
Among many of interesting results concerning the scaling properties of the
BRM model,  very detailed data were presented for the {\it time-averaged}
density-density correlator, known as the ``steady state distribution''
\begin{equation}\label{1}
{\cal K}_t(x_1,x_2)=\overline{\lim\limits_{T\rightarrow \infty }\frac 1T%
\int\nolimits_0^Tdt\left| \langle x_2,t|x_1,0\rangle \right| ^2}=\overline{%
\sum_\alpha \left| \psi _\alpha ^{*}(x_2)\,\psi _\alpha (x_1)\right| ^2},
\label{e:CorrTiAv}
\end{equation}
where $\psi_\alpha $ are eigenfunctions of the Hamiltonian considered to be 
a member of BRM ensemble: $H \psi_\alpha =E_\alpha \psi_\alpha$  and
$E_\alpha $ is the corresponding eigenvalue. The scaled coordinates enetering
Eq.(\ref{1}) are $x_{1,2}=n_{1,2}/b^2$, with $n_{1,2}$ being labels of the 
components of BRM eigenvectors and $b$ being the band widths. The bar stands 
for the averaging over the random matrix ensemble.

In \cite{Izra97} the data for the correlator (\ref{e:CorrTiAv}) has been
fitted by the theoretical expression obtained long time ago by Gogolin%
\cite{Gogo76} for strictly 1D Anderson model. The Gogolin's expression 
contains the localisation length $l$ as a parameter. A reasonable agreement 
has been found for the length parameter value $l_m\approx 0.29 b^2$.  This 
value is rather close to the localization length $l_\infty\big |_{(E=0)}=\frac{1}{3} b^2$
expected  for the {\it orthogonal} BRM ensemble with the matrix size $N \gg  
b^2 \gg 1$ %
\footnote{Note, that in Ref.\cite{Izra97} it was erroneously concluded that
that $l_m$ differs substantially  from $l_\infty$. This is however a  
consequence of incorrect use of the localization length $l_\infty $ defined 
for the {\it unitary} (see Eq.(4),(5) in Ref.\cite{Izra97}) instead of the 
{\it orthogonal} BRM ensemble.  The localisation length $l_\infty$ for the 
unitary class is known to be  twice as large in comparison with that for the 
orthogonal class.)  }%
.

However, the fit done in \cite{Izra97} is in fact not very legitimate. The  
reason is that the Gogolin expression is not relevant for the correlator
(\ref{e:CorrTiAv}) and represents another quantity: a {\it ``fixed energy''} 
correlator defined as
\begin{equation}
{\cal K}_E(x_1,x_2;E)=\overline{\sum\limits_\alpha \left| \psi _\alpha
^{*}(x_2,E_\alpha )\psi _\alpha (x_1,E_\alpha )\right| ^2\delta (E-E_\alpha
).}  \label{e:CorrE}
\end{equation}
It is easy to see, that time-averaged correlator can be obtained from the  
``fixed energy'' one by integration over the energy. Since the localization 
length parameter depends essentialy on the energy,  the integration can modify
the correlator form. Indeed, as we show below the Gogolin correlator with 
energy dependent localization parameter integrated over the energy is a 
better fit for the data. 

Actualy, the best way is to fit the data by theoretical predictions  
following from the BRM model itself, rather than to use Gogolin's formula  
derived formally for diferent case. Unfortunately, the relevant quantity was 
not calculated up to now. In particular, in \cite{Fyod93}
the correlator ${\cal K}_E(x_1,x_2;E)$ has been calculated at points  
$n(x_1)=0$, $n(x_2)=N$ ($N$ is matrix size),  while to compare with  
data\cite{Izra97} one needs such a correlator for finite separations  
$x_2-x_1$ but infinite matrix size $N \to \infty$.

This work is intended to fill this gap. We calculate the density-density  
correlator for infinite banded random matrices of two symmetry classes: the  
{\it orthogonal} ensemble, studied in \cite{Izra97}, and the {\it unitary}  
one, consisting of complex Hermitian banded matrices.  The paper is organized
as follows: after a brief overview of mapping of the BRM model onto the  
supersymmetrical $\sigma$-model in Sec.\ref{BRMdef},  we consider step by 
step the transfer matrix solution of the $\sigma$-model  
in Sec.\ref{TrMatrix} and calculate the {\it fixed energy} correlator in  
Sec.\ref{BRMcorrE} for both symmetry cases simultaneously. As a result,  
we find that within an accuracy of $O(1/b^2)$ both correlators coincide in  
their functional form. At the same time, the localization length is as  
expected twice as smaller in the orthogonal ensemble as compared to the  
unitary one. Integrating in Sec.\ref{BRMcorrT} the ``fixed energy'' 
correlator over the energy we complete the paper with BRM predictions
for the time averaged correlator, the ``steady state distribution''.

\section{Basic definitions of the BRM problem and its mapping onto the  
supersymmetric nonlinear $\sigma$-model.}
\label{BRMdef}

Let us consider the ensemble of $N\times N$ ($N\gg 1$)  banded random  
matrices $H$, whose elements are independently distributed according to 
the Gaussian law with zero mean value and variances
\begin{equation}
\label{e:Hvar}
\overline{\left|H_{ij}\right| ^2}=J_{ij}\equiv f(\left| i-j\right| ),
\end{equation}
The band profile function $f(r)$ vanishes exponentially or faster outside  
the band $\left| i-j\right| \leq b$, $b\gg 1$. In this case, as it was known
from \cite{Fyod91} predictions are insensitive to  the particular band 
profile form.

Below we consider simultaneously two symmetry classes of matrices  $H$: the  
{\it orthogonal} ensemble of real symmeric matrices, and the {\it unitary} 
ensemble  of hermitian ones. The former corresponds to systems which are 
time-reversal invariant , while the latter corresponds to broken invariance.

One can get a solution of the BRM problem  by mapping it onto supersymmetric
$\sigma$-model studied in many detailes in earlier Efetov's  
works\cite{Efet83a,Efet83}.
Let us summarize the results\cite{Fyod91,Fyod94} of such a mapping for both  
symmetry classes. The corresponding  $\sigma$-model action reads:
\begin{equation}
\label{e:Sigma}
S\{Q\} = -\frac{\gamma}{2} \sum_i {\rm Str}Q(i) Q(i+1) +i\epsilon\sum_i {\rm  
Str}Q(i) \Lambda,
\end{equation}
in terms of which the density-density correlator (\ref{e:CorrE}) is expressed
as:
\begin{equation}
\label{e:corrQQ}
{\cal K}(n_1,n_2;E)=-\lim_{\epsilon \to 0} \epsilon\cdot\rho C_0\int \prod_i  
d\mu [Q(i)]\;
(Q^{11}_{bb,11}(n_1)  
Q^{22}_{bb,11}(n_2)+C_\times\delta_{n_1,n_2}Q^{12}_{bb,11}(n_1)  
Q^{21}_{bb,11}(n_1) \exp(-S\{Q\}).
\end{equation}
where the $\sigma$-model supermatrix field $Q$ satisfyes a constraint:  
$Q^2=1$. With use the Efetov's notations the supermatrix $Q$ consists of 
four supermatrix blocks
\begin{equation}
Q=\left(
{\begin{array}{cc}
Q^{11} & Q^{12}\\
Q^{21} & Q^{22}
\end{array}} \right)  ,
\label{e:Q}
\end{equation}
where upper indices 1 and 2 correspond to  the ``retarded'' and ``advanced''  
components of the original RBM model. The bottom indices $bb$ in  
eq.(\ref{e:corrQQ}) denote boson-boson components inside the superblocks. For  
more detailed explanation see, e.g. reviews \cite{Verb85,Fyod94}.

The difference between the two symmetry classes is reduced to  
symmetry-specific definitions of the model parameters 
$\rho,\; \gamma ,\; C_0,C_\times$ %
\footnote{Note, that the factor $C_\times$ originates from the number of 
cross-pairing of components $1\leftrightarrow 2$. Since this contribution 
gives rise at $n_1=n_2$ only, it is of main importance in inverse 
participation ratio (IPR).}%
, as well as of the supermatrices $\Lambda$ and $Q$.

In the {\it orthogonal} ensemble the $8\times8$ supermatrices Q belong  to   
the  coset space  UOSP(2,2/2,2)/ UOSP(2/2)$\times$UOSP(2/2).
Their explicit parametrization, including invariant measure $d\mu[Q]$  can  
be found, e.g. in \cite{Verb85,Verb88}.
The diagonal matrix $\Lambda$ is $\Lambda={\rm diag}(1,1,1,1,-1,-1,-1,-1)$. 
The density of states $\rho$ obeys the semicircle law
\begin{equation}
\rho=\frac{1}{\pi J_0 } (E^2_{max} - E^2)^{1/2}, \hskip 2em  \left|  
E\right|\leq E_{max}=(2 J_0)^{1/2},
\label{e:DOS_GOE}
\end{equation}
and the ``length scale'' parameter $\gamma$ is  \cite{Fyod92}
\begin{equation}
\gamma=\frac{(\pi \rho J_0)^2}{4} B=\frac{1}{4} B (2 J_0-E^2),
\label{e:gammaGOE}
\end{equation}
where the band profile parameters are defined as
\begin{equation}
J_0=\sum^{\infty}_{r= -\infty} a(r) , \hskip 2em  B=\sum^{\infty}_{r=  
-\infty} a(r) r^2/(J_0)^2.
\label{e:JoB}
\end{equation}
The correlator parameters are $C_0=2$ and $C_\times=2$, correspondingly..

In the {\it unitary} ensemble the $4\times4$ supermatrices Q belong  to  the  
coset space U(1,1/2)/U(1/1)$\times$U(1/1),
and their explicit parametrization can be found, e.g. in \cite{Zirn86,Fyod94}.
In this case the matrix $\Lambda$ is $\Lambda={\rm diag}(1,1,-1,-1)$. The  
density of states $\rho$ as well as the parameter $\gamma$ are given by  
\cite{Fyod94}:
\begin{equation}
\label{e:DOS_GUE}
\rho=\frac{1}{2\pi J_0 } (E_{max}^2 - E^2)^{1/2}, \hskip 2em \left|  
E\right|\leq E_{max}= (4 J_0)^{1/2},
\end{equation}
\begin{equation}
\label{e:gammaGUE}
\gamma=(\pi \rho J_0)^2 B=\frac{1}{4} B (4 J_0-E^2),
\end{equation}
and differ from those in the orthogonal ensemble. It is easy to see, that in  
the unitary case the ``length scale'' $\gamma$ (at $E=0$)
is twice as large and the correlator parameters are in this case: $C_0=1$,  
$C_\times=1$.

At this level of consideration two symmetry classes differ mainly by the  
rescaling of the length scale $\gamma$ and by explicit form of the 
supermatrices $Q$.

At the end of this section let us consider briefly the accuracy of the BRM  
model mapping onto the supersymmetrical $\sigma$-model. The only  
approximation done at this step is the integration over so-called ``massive''  
modes performed by saddle-point method in Gaussian approximation(for more  
details see, e.g. \cite{Verb85,Fyod94}). While the separation between
the saddle points is of order (see, e.g. \cite{Verb85,Fyod94})
\begin{equation}
  \Delta\sim\pi\rho J_0 \sim J_0^{1/2}(E^2_{max}-E^2)/E^2_{max},
\end{equation}
the variances of the massive modes are $\overline{(\delta P)^2}\propto  
J_0/b$ \cite{Fyod94}. Hence, the ratio of the amplitude of the massive modes  
to the sadle points separation is of order of $O(b^{1/2})$:
\begin{equation}
   r\sim \frac{\left(\overline{(\delta P)^2}\right)^{1/2}}{\Delta}\sim  
\frac{E^2_{max}}{b^{1/2}(E^2_{max}-E^2)^{1/2}}
   \sim \left(\frac{E_{max}}{b(E_{max}-|E|)}\right)^{1/2},
\end{equation}
 which, probably, defines the accuracy of the approxiamtion.%
\footnote{Since the number of saddle is, in fact, even, the cancellation of  
terms linear in $r$ can occur.  Then the accuracy will be better, of order of  
$O(b^{-1})$. More careful consideration is needed to arrive at definite 
conclusion.} 
Putting $r\sim 1$, one may also conclude,
that semicircle law for the density of states is valid as long as $\Delta  
E=(E_{max}-|E|)\gg 1/b$.

\section{Transfer matrix solution: from the discrete model to the continuous 
one.}
\label{TrMatrix}

Starting with the pioneer work\cite{Efet83a}, the transfer matrix technique  
looks as the most adequate tool to solve the problem. Forthcoming 
works\cite{Zirn86,Verb88} make some detailes be more transparent. The 
continuous limit version \cite{Fyod92} reduces the problem to solution of a 
partial differential equation. As we check below, this equation
is the same for both the orthogonal and unitary ensemble, with accuracy  
$O(\gamma^{-1})$. The corresponding boundary conditions are also the same 
when calculating the density-density correlator, but can be symmetry dependent
in more general case. To demonstrate this let us outline basic steps of the 
approach.

Following \cite{Fyod94} consider the N-site one-dimensional chain of  
matrices $\{ Q(k)\}$, with a nearest neighbour interaction given by 
eq.(\ref{e:Sigma}). For the sake of definiteness we assume that  $n_1<n_2$ in 
eq.(\ref{e:corrQQ}). Having in mind that the integrations over different 
$Q(k)$, $k=1,...(n_1-1)$ in eq.(\ref{e:corrQQ}) can be performed subsequently 
one after another with the {\it same} kernel
\begin{equation}
\label{e:L_QQ}
  L(Q,Q')=\exp\{\frac{\gamma}{2} {\rm Str}Q Q' - i\epsilon {\rm Str} Q'\Lambda \},
\end{equation}
it is very convenient to consider the quantity $Y^{(1)}(Q; k)$ satisfying  
the recurrence relation
\begin{equation}
\label{e:LYqq}
 Y^{(1)}(Q;k)=\int \; d\mu(Q') L(Q,Q') Y^{(1)}(Q';k-1).
\end{equation}
with the initial condition $Y^{(1)}(Q;0)\equiv 1$. In general, the integration
over $Q$ is quite nontrivial. A great simplification occurs in the limit 
$\epsilon\to 0$, which is exactly the case we are interested in (see the 
eq.(\ref{e:corrQQ})). As it was clearly demonstrated in\cite{Zirn86,Verb88} 
with Efetov parametrization of matrices $Q$, after taking such a limit the 
function $Y^{(1)}(Q;k)$ depends only on one of its bosonic ``eigenvalues''
(say, $\lambda_1$) of the superblock $Q^{11}$, which takes values  
$\lambda_1\sim \epsilon^{-1}$. Although the kernel (\ref{e:L_QQ}) has a very 
extensive grassmannian part, in the limit $\epsilon\to 0$ only the term of 
the maximal grassmanian order survives due to the Parisi-Sourlas-Efetov-Wegner
(PSEW) theorem%
\footnote{See, for details \cite{Zirn86,Cons88}.}. This is the term of the  
eighth or fourth order in the orthogonal and unitary case, respectively.
Perfoming explicit integration over all variables but $\lambda_1$ and using,  
that $Y^{(1)}(Q;k)$ depends only on $z=2\epsilon \lambda_1$, one can rewrite 
the recurrence relation (\ref{e:LYqq}) in a more simple form:
\begin{equation}
  Y^{(1)}(z;k)=\int_0^\infty dz' \; L_\gamma(\frac{z}{z'})\exp(-z) Y^{(1)}(z';k-1).
\label{e:LYz}
\end{equation}
The explicit form of the kernel $L_\gamma(\frac{z}{z'})$ can be found, e.g.  
in \cite{Verb88} and \cite{Zirn86} for the orthogonal and unitary case, 
respectively.

This recurrence relation can be used iteratively to construct the function  
$Y^{(1)}(z;k)$ up to the point $k=n_1$. At this point the integrand
in eq.(\ref{e:corrQQ}) has an extra factor $Q^{11}_{bb,11}$. This factor  
already  depends on the whole set of the ordinary parameters of  matrix $Q$  
and on a half subset of grassmann variables (the two $\alpha,\alpha^*$ in 
the unitary case, and the four $\alpha_1,\alpha^*_1,\alpha_2,\alpha^*_2$
in the orthogonal case, in notations of papers\cite{Zirn86,Verb88})
\footnote{Analogous factor $Q^{22}_{bb,11}$, with replacing $\beta \to  
\alpha$, arises at $k=n_2$.}.
Omitting the detailes ( some of them can be found in \cite{Verb88,Fyod94})  
we have found that in the limit $\epsilon\to 0$ the only nonvanishing 
contribution to the density-density correlator  coming from the factor 
$Q^{11}_{bb,11}$ is the  term of the maximal order in grassmann variables. 
It is of the kind $c(z,...)G_{max}(\alpha)$, with
\begin{equation}
G_{max}(\alpha)\equiv
\left\{\begin{array}{ll} \alpha_1\alpha^*_1\alpha_2\alpha^*_2, \hskip 1em&  
{\rm orthogonal\;\;case,}\\
           \alpha\alpha^*, & {\rm unitary\;\; case,}
\end{array}\right.
\end{equation}
and the ordinary coefficient  $c(z,...)$, which (after averaging over the  
angle variables in the orthogonal case) is the same for both symmetry  
classes: $c(z,...)\to(-iz/(2\epsilon))$ .

To proceed further, beginning from the point $k= n_1$ we start iterations  
with a new function $Y_G^{(2)}(Q; k, n_1)$ satisfying initial condition
$Y_G^{(2)}(Q; k=n_1,n_1)=zY^{(1)}(z,n_1) G_{max}(\alpha)$. The subscript $G$  
denotes here, that this new function aquires a {\it nontrivial grassmann part}.
As a result, one could suspect that the recurrence relation (\ref{e:LYz}) 
is not valid any longer and one should return to the more complex relation  
(\ref{e:LYqq}). Fortunately, the closer inspection shows that to the leading 
order in $\epsilon\to 0$ the recurrence relation (\ref{e:LYqq}) acts on 
functions $Y_G^{(2)}(Q; k, n_1) =Y^{(2)}(z;k,n_1)\cdot G_{max}(\alpha)$  
in such a way as if the grassmannian factor remains unchanged, while its 
{\it ordinary}  coefficient  $Y^{(2)}(z;k,n_1)$ is iterated by the relation 
(\ref{e:LYz}) with {\it the same} kernel, as before.  

The functions $Y^{(1)}(z; k)$ and $Y^{(2)}(z; k, n_1)$ allow one to  
calculate the correlator (\ref{e:corrQQ}). Let us first consider
the ``direct pairing'' contribution to eq.(\ref{e:corrQQ}) coming from the  
term $Q^{11}_{bb,11}(k) Q^{22}_{bb,11}(l)$:
\begin{equation}
\label{e:corrQQ1122}
{\cal K}_d(n_1,n_2;E)=-\lim_{\epsilon \to 0} \cdot\epsilon\rho C_0
\int  d\mu (Q) \; Y^{(1)}(z=2\epsilon\lambda_1; N-n_2)\;Q^{22}_{bb,11}  
g_{max}(\alpha)\;(-\frac{i}{2\epsilon})
Y^{(2)}(z=2\epsilon\lambda_1; n_2,n_1) e^{-z},
\end{equation}
where the only explicit integration over $Q$ left is that done at the site  
$k=n_2$. Using the PSEW theorem, performing all the integrations%
\footnote{Note, that this integrations give the factor 1 and 2 in the  
orthogonal and unitary cases, respectively. With
the factor $C_0$ this gives a factor 2 for the both symmetry cases.}%
except for that over $z=2\epsilon\lambda_1$, and finally taking the 
limit $\epsilon\to 0$ we come to the expression
\begin{equation}
\label{e:corrQ11Q22}
{\cal K}^{(d)}_E(n_1,n_2;E)=  \rho\int  \frac{dz}{z^2} \; Y^{(1)}(z;  
N-n_2)\;z\;Y^{(2)}(z; n_2,n_1) e^{-z}.
\end{equation}
Note, that this expression is the same for  both symmetry classes, but the  
functions $Y^{(1)}(z; n_2)$ and $Y^{(2)}(z; n_2, n_1)$
are calculated with kernels $L_\gamma$, which, generally speaking, are {\it  
dependent} on the symmetry class.

Another contribution to the correlator (\ref{e:corrQQ})comes from the  
cross-pairing term $Q^{12}_{bb,11} Q^{21}_{bb,11}$. It is
even more easy to calculate. One can verify, that at the point $n_2=n_1$   
the term $Q^{12}_{bb,11} Q^{21}_{bb,11}\approx Q^{22}_{bb,11} Q^{11}_{bb,11}$  
to the leading order in $\lambda_1\sim 1/\epsilon\to \infty$.
Therefore the full correlator can be written as
\begin{equation}
{\cal K}_E(n_1,n_2;E)={\cal K}^{(d)}_E(n_1,n_2;E)+C_\times\delta_{n_1 n_2}  
{\cal K}^{(d)}_E(n_1,n_1;E).
\end{equation}

The final simplification arises if one considers the bandwidth $b$ of the  
matrices to be large enough, so that the parameter $\gamma\sim b^2 \gg 1$. 
Let us note, that this condition was actually used in the previous section 
in order to map the original BRM problem onto the sypersymmetrical 
$\sigma$-model (see discussion in \cite{Fyod94}).

First of all,  the kernels $L_\gamma(z/z')$ obtained in \cite{Verb88} for 
the orthogonal case as well as that for the unitary case \cite{Zirn86}  
coincide with each other within the accuracy $O(\gamma^{-1})$. Then one can 
pass to the continuous limit by using that the kernel has a sharp peak. 
It amounts to replace the integral recurrence equation by the differential  
one\cite{Fyod92,Fyod94}:
\begin{equation}
\label{e:Diff}
\frac{\partial W^{(s)}(y,x)}{\partial \tau}=\hat{\bf R} W^{(s)}(y,x), \hskip 2em 
\hat{\bf R}\equiv\left\{y^2\frac{d^2}{dy^2}-y\right\}, \hskip 2em s=1,2;
\end{equation}
where we introduced a continuous variable $x=k/2\gamma$, a scaled variable  
$y=2\gamma z$, and the functions
\begin{eqnarray}
\label{e:WW}
&&W^{(1)}(y;x)= Y^{(1)}(z=y/2\gamma; k=2\gamma x ), \\
&&W^{(2)}(y;x,x_1)= 2\gamma\cdot Y^{(2)}(z=y/2\gamma; k=2\gamma x,  
n_1=2\gamma x_1);
\end{eqnarray}
satisfying the initial conditions
\begin{equation}
\label{e:WWin}
W^{(1)}(y;0)=1, \hskip 2em W^{(2)}(y;x_1,x_1)=yW^{(1)}(y;x_1).
\end{equation}
Let us stress, that the differential equation (\ref{e:Diff}) as well as the  
initial conditions (\ref{e:WWin})
are exactly the same for both  symmetry classes%
\footnote{Note, that in the course of calculations of other quantities, e.g.  
{\it the probability distribution} for the
density-density correlator or inverse participation ratio, the corresponding  
initial conditions may be
{\it symmetry dependent}\cite{Fyod93}.}.

The correlator (\ref{e:corrQQ1122}) in continuous limit takes a form
\begin{equation}
\label{e:CorrW}
{\cal K}^{(d)}_E(x_2,x_1;E)= \frac{\rho}{2\gamma}\int  \frac{dy}{y^2} \;  
W^{(1)}(y; x_N-x_2)\;y\;W^{(2)}(y; x_2,x_1) e^{-y/2\gamma},
\end{equation}
where we introduced the dimensionless matrix size as $x_N=N/2\gamma$.

\section{Calculation of the correlator ${\cal K}_E(x_1,x_2;E)$ for infinite  
banded matrices.}
\label{BRMcorrE}

The way to obtain a solution of eq.(\ref{e:WW}) was in details considered in  
\cite{Fyod92,Fyod94}. For the function $W^{(1)}(y,x,x_0)$ one has
\begin{equation}
W^{(1)}(y;x) =2y^{1/2}\left\{ K_1(2y^{1/2})+\frac 2\pi
\int\limits_0^\infty d\nu _1\frac{\nu _1}{1+\nu _1^2}\sinh (\pi \nu
_1)K_{iv_1}(2y^{1/2})\,e^{-\frac{1+v_1^2}4x}\right\},
\label{e:W1}
\end{equation}
where $K_\alpha (t)$ is the MacDonald function. In this work we consider the  
case of the band matrices of the infinite size:  $N/2\gamma \to \infty$.  In  
contrast to \cite{Fyod93,Fyod94} the correlator arguments $x_1,x_2$ are  
taken at the finite separation $(x_2-x_1)$, but both far from the ends 
$x=0$, $x_N=N/2\gamma$. In this limit we have
\begin{equation}
W^{(1)}(y;x_1),W^{(1)}(y;x_N-x_2) \rightarrow 2y^{1/2}K_1(2y^{1/2})%
\label{e:W1inf}
\end{equation}
 since  both
$x_1 \to \infty$, and $(x_N-x_2)\to\infty$.
Constructing the function $W^{(2)}(y,x,x_1)$ with the initial condition 
(\ref{e:WWin}): $W^{(2)}(y,x_1,x_1)=y\cdot 2y^{1/2}K_1(2y^{1/2})$ we find
\begin{equation}
W^{(2)}(y;x_1,x_2)=\frac{y^{1/2}}{\pi ^2}\int\limits_0^\infty d\nu _2\,\nu
_2\sinh (\pi \nu _2)\,e^{-\frac{(1+\nu _2^2)}4(x_2-x_1)}K_{i\nu
_2}(2y^{1/2})\int\limits_0^\infty d\tilde{t}\,\tilde{t}K_{i\nu _2}(%
\tilde{t}).  
\label{e:W2}
\end{equation}
Substituting the functions $W^{(1)}$ and $W^{(2)}$ into the correlator  
expression (\ref{e:CorrW}) one finally arrives at
\begin{eqnarray}
\frac{1}{\rho}{\cal K}^{(d)}_E(x_2,x_1;E) &=&\frac 1{2\gamma }\int\limits_0^\infty
dy\,y^{-1}e^{-\frac{y}{2\gamma }}\cdot 2y^{1/2}K_1(2y^{1/2})\times   \nonumber
\\
&&\hskip 2em\times \frac{y^{1/2}}{\pi ^2}\int\limits_0^\infty d\nu _2\,\nu  
_2\sinh(\pi \nu _2) \,e^{-\frac{(1+\nu _2^2)}{4}(x_2-x_1)}  
K_{i\nu_2}(2y^{1/2})\int\limits_0^\infty d\tilde{t}\,\tilde{t}K_{i\nu  
_2}(\tilde{t})  \label{e:Corr1}%
\\
&\simeq &\frac 1{2\pi ^2\gamma }\int\limits_0^\infty d\nu _2\,\nu  
_2\sinh(\pi \nu _2)\,e^{-\frac{(1+\nu _2^2)}4(x_2-x_1)} \left[  
\int\limits_0^\infty dt\,tK_1(t)K_{iv_2}(t)\right] ^2  \nonumber
\\
&=&\frac{\pi ^2}{16\cdot 2\gamma }.\int\limits_0^\infty d\nu \,\nu \sinh(\pi  
\nu )\,e^{-\frac{(1+\nu ^2)}4(x_2-x_1)} \left( \frac{1+\nu ^2}{1+\cosh\pi  
\nu }\right) ^2 \label{e:Corr2}
\end{eqnarray}
In  eq.(\ref{e:Corr1}) we omitted the factor $\exp (-y/2\gamma )$  since the  
integral over $y$ converges at $y\sim 1$ but $2\gamma \sim b^2\gg 1$. We see  
also, that in the infinite matrix size limit the correlator depends on the  
difference $x=x_2-x_1$ only which means the restoration of the translational  
symmetry.

It is interesting, that the BRM model prediction for the {\it functional  
form} of the correlator ${\cal K}_E(x_2-x_1;E)$ coinsides exactly with that  
obtained many years ago by Gogolin \cite{Gogo76} for the strictly 1D electron 
system in disordered potential. However, the spatial scales in these systems 
are quite different. Namely, in the {\it strictly} 1D electron system it is equal to 
the mean free path defined by the Born amplitude of electron-impurity 
scattering, while in the {\it quasi}-1D BRM model it is governed by the 
localisation length $l_c=2\gamma\propto b^2$, growing quadratically 
with matrix bandwidth.

\section{RBM model predictions for time-average correlator ${\cal K}_t(x_1,x_2)$.}
\label{BRMcorrT}

As we have already mentioned in the Introduction, the {\it time-averaged}  
correlator ${\cal K}_t(x_1,x_2)$ can be obtained from the ``fixed energy'' 
correlator ${\cal K}_E(x_1,x_2;E)$ by integrating the latter over the energy:
\begin{equation}
{\cal K}_t(x_1,x_2)=\int\limits_{-\infty }^\infty dE\,{\cal K}_E(x_1,x_2;E).
\label{e:QErel}
\end{equation}
Substituting eq.(\ref{e:Corr2}) into eq.(\ref{e:QErel}) we get
\begin{equation}
{\cal K}_t(x_2-x_1)=\int\limits_{-E_0}^{E_0}dE\,\frac{\pi ^2\rho  
(E)}{16\cdot l_c(E)}.\int\limits_0^\infty d\nu \,
\nu \sinh (\pi \nu )\,\cdot 
 \exp\left\{-\frac{(1+\nu ^2)}{4}\frac{l_c(0)}{l_c(E)}(x_2-x_1)\right\}
\left(\frac{1+\nu ^2}{1+\cosh \pi \nu}\right) ^2,  
\label{e:K_t}
\end{equation}
with energy dependent localization length $l_c(E)=2\gamma (E)$. The scaled
variables $x_{1,2}$ are related here to the discrete ones $n_{1,2}$ via
the energy independent scale factor $l_c(0)$: $x_{1,2}=n_{1,2}/l_c(0)$. 

For large $N,b\gg 1$ the density of states $\rho (E)$ in RBM model obeys  
the semicircle law:
\begin{equation}
\rho (E)=\rho (0)\cdot \left( 1-\frac{E^2}{E_0^2}\right) ^{1/2}=\frac 2{\pi
E_0}\left( 1-\frac{E^2}{E_0^2}\right) ^{1/2},  \hskip 2em  \int dE\,\rho  
(E)=1;\label{e:DOS(E)}
\end{equation}
and the localization length is, correspondingly,
\begin{equation}
l_c(E)=l_c(0)\cdot \left( 1-\frac{E^2}{E_0^2}\right)
\label{e:Lc(E)}
\end{equation}
with $E_0=2J_0, \; l_0(0)=J_0B/2\approx b^2/3$  or $E_0=4J_0, \;  
l_0(0)=BJ_0\approx 2b^2/3$ for the orthogonal or unitary cases, respectively  
(see eqs.(\ref{e:DOS_GOE})-(\ref{e:gammaGUE})).

Let us make a remark about the limits of the integration. According to the 
note at the end of Sec.\ref{BRMdef}, both the semicircle law for the density 
of states and the solution itself are valid as long as $(E_0-|E|) \gg 
\delta E\sim E_0/b$. Thus, extending the integration in the Eq.(\ref{e:K_t})
over the whole region $[-E_0,E_0]$ we get an uncertainty coming from the
regions $E_0-|E|\leq \delta E$, where our integrand differs substantially
from the exact one. To estimate an order of magnitude of this uncertainty, 
let us consider the contribution coming from this regions with {\it our} 
integrand. In the ``worst'' case of the coinciding correlator arguments 
$x_1=x_2$ the integrand has integrable {\it square root} singularities 
at the ends $E=\pm E_{max}$, and their contribution has a relative order
of magnitude $(\delta E/E_0)^{1/2}\sim b^{-1/2}$. In the case of nonzero
$x=x_2-x_1$ the contribution of of the regions $E_0-|E|\leq \delta E$
is exponentially small at $x\gg b^{-1}$. In particular, we conclude
that in the recent numerical BRM simulations\cite{Izra97} done for the matrix 
bandwidth $b=4-12$ substantial deviations order of $b^{-1/2}$ can occur 
for the correlator at $x\lesssim b^{-1}$.

The integration in eq.(\ref{e:K_t}) can hardly be performed analytically.  
Let us therefore consider  two simple limiting cases first.
It can be most easily done for the correlator value at $x=(x_2-x_1)=0$:
\begin{equation}
{\cal K}_t(0)=2{\cal K}_E(0;E)\mid _{E=0}=\frac{2}{3l_c(0)}(1+C_\times),   
\label{e:K_t0}
\end{equation}
which is a ``time-averaged'' version of the so-called inverse participation  
ratio (IPR) $\xi_t\equiv
\sum_\alpha \left| \varphi _\alpha \right| ^4$. Note, that {\it in the BRM  
model} ``time-averaged'' IPR is twice as large as compared to its ``fixed  
energy`` counterpart  usually considered in the solid physics and defined as  
$\xi _E\equiv \sum_\alpha \left|
\varphi _\alpha \right| ^4\delta (E_\alpha -E)$ for quantum states near  
$E=0$. It is a natural consequence of the  averaging over localized quantum  
states with $l_c(E)<l_0(0)$.

Large-$x$ asymptotics $(x\gg l_c(0))$ can be calculated by the  
steepest-descent method. Performing the calculation, we find
\begin{equation}
{\cal K}_t(x)=\left[ \frac 4{\sqrt{\pi }}\left( \frac{l_c(0)}{\left|
x\right| }\right) ^{1/2}\right] {\cal K}_E(x;E)\mid _{E=0}.  \label{e:K_tx}
\end{equation}
Comparing with the expression eq.(\ref{e:K_t0})we see that at large $x$ the  
extra factor reduces the value of the correlator. Therefore, converting  
${\cal K}_E$ into ${\cal K}_t$ {\it changes} the correlator {\it form}, most  
considerably in the region $\left| x\right|\lesssim l_c(0)$.

In the recent numerical work \cite{Izra97} the data for the correlator  
${\cal K}_t(x)$ have been obtained in the course of a direct numerical  
simulation of the BRM orthogonal ensemble. With use of the Gogolin  
formula\cite{Gogo76} the best fit value for the localisation length was found  
$l_c\approx 0.29 b^2$. In Fig.1 we plot the correlator ${\cal K}_t(x)$ 
predicted by eq.(\ref{e:K_t}) which has {\it no free parameters} and 
corresponds to $l_c(E=0)=b^2/3$
\footnote{To facilitate the comparison with data we plot
the curves in terms of scaled variable $x=n/b^2$, as defined in 
Ref.\cite{Izra97}}%
.  The dashed line presents the parametrization
of the data by the Gogolin formula with $l_c= 0.29 b^2$.
To illustrate an accuracy of the agreement, we plot also the Gogolin formula  
with $l_c=  b^2/3$. We see, that the agreement looks reasonable. For more 
reliable analysis one needs to  compare the correlator ${\cal K}_t(x)$ 
directly with the data for BRM ensembles with the matrix bandwith $b\gg 1$.

However, the most crucial test of our theory would be the comparison of the  
correlator behavior at $x\lesssim 1$ with numerical data (see, Fig.2). We 
expect it to be essentially more peaked at $x=0$ than the ``fixed energy'' 
correlator  ${\cal K}_E(x_2-x_1;E)$ given by the Gogolin formula with 
$l_c=b^2/3$. Apart from this fact, it would also be interesting to compare 
the {\it time-averaged} IPR with data, which is predicted to be twice as 
large in comparison with the corresponding ``fixed energy'' quantity.

\section{Conclusions.}
\label{Concls}

In this work we have studied the density-density correlators in the BRM  
ensembles for the orthogonal and unitary symmetry cases. In contrast to the 
works\cite{Fyod94,Fyod91,Fyod93,Fyod92} we consider the case of infinite 
matrix size $N\to\infty$,  with correlator arguments $n_1\leq n_2$ taken 
well inside the argument region: $[1,...,n_1,...,n_2,...,N]$, 
far from the ends: $n_1\gg b^2$,  $N-n_2\gg b^2$ ($b$ is matrix bandwidth).

We have found, that the ``fixed energy'' density-density correlator  
(\ref{e:CorrE}) turns out to be the same in both the orthogonal and unitary  
BRM ensembles, within the accuracy $O(b^{-2})$. Moreover, its functional form  
coincides exactly with the old Gogolin formula originally derived for 1D  
electron system in  disordered potential.

It deserves to be mentioned that the analytical solution of the BRM model  
given in \cite{Fyod91,Fyod94} consists of several approximation steps. 
The first one is  the mapping of the model onto the supersymmetrical 
$\sigma$-model. It has, probably,  the worst accuracy of order $O(b^{-1/2})$. 
At the same time the accuracy of the $\sigma$-model solution, including 
the continuous approximation, is the order $O(\gamma^{-1}) \sim O(b^{-2})$. 
Thus, one may expect deviations from the exact solution to be of order of 
$O(b^{-1/2})$.

Beside the ``fixed energy'' density-density correlator ${\cal K}_E(x_2-x_1;E)$,
we have also calculated the {\it time-averaged} correlator   
${\cal K}_t(x_2-x_1)$. Two functions are found to be different, especially  
in the region of $(x_2-x_1)\lesssim 1$. We predict also, that the 
{\it time-averaged} inverse participation ratio for BRM model is twice as 
large as the ``fixed energy'' one.

\section{Acknowledgements}

I am very grateful to B.V. Chirikov, who stimulate my attention to this  
problem, for the support, numerouos useful considerations and critical  
remarks, and to Y.V. Fyodorov for critical reading of the manuscript and 
invaluable help on its preparation for publication. 
I am also very thankful to V.V.  Sokolov, D.V. Savin and I.V. Kolokolov 
for the useful conversations and discussion of the results.

\begin{figure}
\caption{The {\it time-averaged} density-density correlator ${\cal K}_t(x)$ versus $x=n/b^2$. The solid line represents
the prediction by eq.(\ref{e:K_t}). The dashed line shows the parametrization of the numerical data[3], the Gogolin
formula with $l_c=0.29$, and dotted line corresponds the same Gogolin formula with $l_c=0.33$.}
\end{figure}

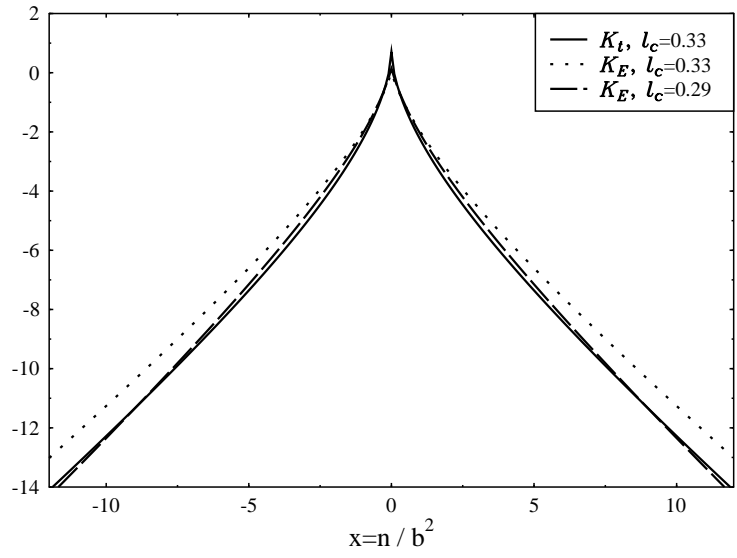
\begin{figure}
\caption{The {\it time-averaged} density-density correlator ${\cal K}_t(x)$ versus $x=n/b^2$ (central region). The solid line represents prediction by eq.(\ref{e:K_t}). The dotted line shows  the Gogolin formula with $l_c=0.33$.}
\end{figure}


\begin{references}
\bibitem{Izra90}  F.M. Izrailev, Phys. Rep. {\bf 196} (1990) 299.

\bibitem{Fyod94}  Y.V. Fyodorov and A.D.Mirlin, Int. Journ. Mod. Phys. B
{\bf 8} (1994) 3795.

\bibitem{Izra97}  F.M. Izrailev, T. Kottos, A. Politi and G.P.Tsironis,
Phys. Rev. E {\bf 55}, (1997) 4951.

\bibitem{Guhr97} T.Guhr, A. M\"{u}ller-Groeling and H.A. Weidenm\"{u}ller,
cond-mat/9707301.

\bibitem{Fyod91}  Y.V. Fyodorov and A.D.Mirlin, Phys. Rev. Lett. {\bf 67}
(1991) 2405.

\bibitem{Efet83}  K.B. Efetov, Adv. Phys. {\bf 32} (1983) 53.

\bibitem{Casa90} G.Casati, L. Molinari, and F.M. Izrailev, Phys. Rev Lett.  
{\bf 64}, 1851 (1990);
      S.N. Evangelou E.N. Economu, Phys. Lett. A {\bf 151}, 345 (1990);  M.  
Feingold, D.M. Leitner, and
      M. Wilkinson, Phys. Rev. Lett. {\bf 66}, 986 (1991).

\bibitem{Gogo76}  A.A. Gogolin, ZhETP {\bf 71} (1976) 1912.

\bibitem{Fyod93}   Y.V. Fyodorov and A.D.Mirlin, Pis'ma v Zh. Eksp. Teor.  
Fiz. {\bf 58}, 636 (1993) [JETP Lett. {\bf 58},
      615 (1993)].

\bibitem{Efet83a} K.B. Efetov and A.I. Larkin, Zh. Eksp. Teor. Fiz. {\bf 85}  
764 (1983) [Sov. Phys. JETP {\bf 58}, 606
      (1983)].

\bibitem{Verb85} J.J.M. Verbaarschot, H.A. Weidenm\"{u}ller, and  
M.R.Zirnbauer,  Phys. Rep. {\bf 129}, 367 (1985).

\bibitem{Verb88} J.J.M. Verbaarschot, Nucl. Phys. B {\bf 300}, 263 (1988).

\bibitem{Fyod92}  Y.V. Fyodorov and A.D.Mirlin, Phys. Rev. Lett. {\bf 69}  
1094 (1992);

\bibitem{Zirn86} M.R.Zirnbauer, Nucl. Phys. B {\bf 265}, 375 (1986).

\bibitem{Cons88} F. Constantinescu and H.F. de Groote, J. Math. Phys. {\bf  
30} 981 (1988).


\end{references}
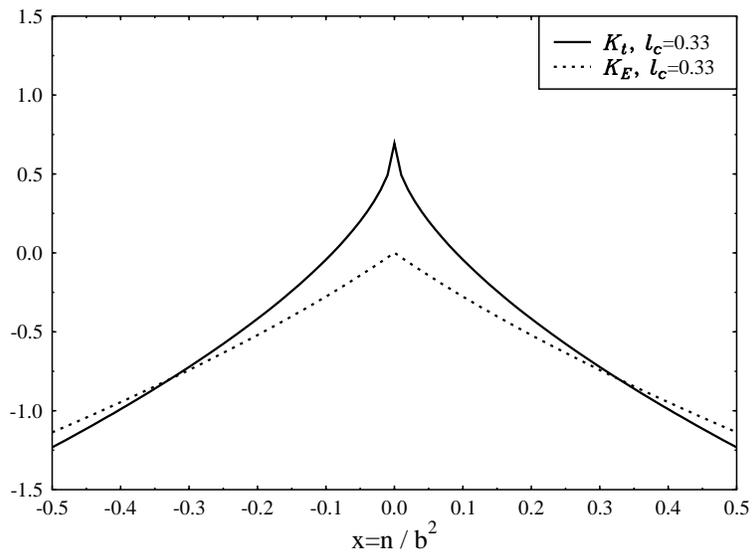
\end{document}